# Task queue implementation for edge computing platform

Veljko Maksimović[1][0000-0002-4905-2421], Miloš Simić[1][0000-0001-8646-1569], Milan Stojkov[1][0000-0002-0602-0606], and Miroslav Zarić[1][0000-0003-4671-3834]

[1] University of Novi Sad, Faculty of Technical Sciences, Novi Sad, Serbia
`{veljko.maksimovic, milos.simic, stojkovm, miroslavzaric}@uns.ac.rs`

**Abstract.** With the rising number of distributed computer systems, from microservice web applications to IoT platforms, the question of reliable communication between different parts of the aforementioned systems is becoming increasingly important. As part of this paper, a task queue, which facilitates reliable asynchronous communication between different services, will be implemented. In order to control the flow of tasks through the queue and limit the load on downstream components, we are going to explore ways of efficiently restricting throughput, and defining priority queues within the task queue service. The research will also take a look at how different aspects of the task queue, such as the underlying persistence layer, affect their performance, reliability, and resource usage. This task queue will be implemented as a component in the already existing platform for managing clusters, *constellations[1]*.

## 1   Introduction

### 1.1   Cloud computing

At the beginning of the 1960s, a new paradigm for developing and running software applications emerged - cloud computing. The idea was to share computer resources, such as RAM, CPU, and hard disk space. Developers would not have to worry about assembling, monitoring, and repairing hardware components, and could focus on software. Later on, the distribution of entire platforms such as operating systems, web servers, and databases was introduced [1].

Running applications in these cloud environments has a few drawbacks. Data centers where applications are hosted can be distant from end users, introducing delays due to network communication [2]. There are also scenarios, depending on the countries where the application is being used, and the domain, in which user data can't leave state borders [3]. In this case, applications need to be hosted in each state individually, which could eliminate some cloud providers who have their data centers in only a handful of locations.

---

[1] https://github.com/c12s

## 1.2 Edge computing

"Edge computing is a new computing model that deploys computing and storage resources (such as cloudlets, micro data centers, or fog nodes, etc.) at the edge of the network closer to mobile devices or sensors." [4] It adds another layer of infrastructure to modern information systems that is geographically closer to end users. The goal is the reduction of traffic between clients and data centers by offloading some of the tasks to the edge nodes. Internet service providers (ISPs) have been using servers on the edge of their networks for caching web pages so they wouldn't have to route all of the requests to their final destination, therefore reducing communication costs [5].

A similar approach could be applied to cloud computing: The entire infrastructure can be expanded by adding smaller clusters of compute nodes near every bigger population center. Developers would be able to run processes for aggregating or anonymizing data on those nodes before passing it on to the central cluster for further processing and storage. These small clusters positioned geographically near users are called *micro-clouds* [6].

The goal of the *constellations*[2] project, explained in detail in *Towards Edge Computing as a Service: Dynamic Formation of the Micro Data-Centers* is to create a control unit for provisioning and managing *micro-clouds*. Further implementation of the task queue, which this paper follows, will be done as a part of the above-mentioned project.

## 1.3 Current Research

Edge computing introduces a new way to structure applications, but it also introduces some new challenges. Since the hardware resources that are available at the edge aren't as powerful as those in the cloud, some of the modern security best practices like complex cryptography algorithms aret feasible. Also, this is a computing mode that integrates multiple trust domains with authorized entities as trust centers, traditional data encryption and sharing strategies are no longer applicable. [7] Currently, research into security and privacy protection for edge computing is still in the early stages, with limited findings available.

Another area of research is computation offloading. Given the constraints in computational power mentioned above, it's critical to use these resources efficiently. Different strategies for offloading are being researched, where decisions are made on whether to process data locally or send it to a more powerful node or the cloud. This takes into account aspects like workloads, required response times, and the state of the network. [8] Machine learning and AI are also being employed to predict and adapt to varying edge environments, ensuring optimal performance and minimal resource wastage.

---

[2] https://github.com/c12s

## 2       Communication in distributed systems

This chapter will briefly go over different ways of implementing communication over the network between different services.

### 2.1    **Synchronous communication**

Synchronous communication between software components implies that the component that initiated communication is waiting for the response in the idle state, before executing any other commands. This isn't a problem if both components are part of the same process, or different processes on the same operating system, as the time to pass the message from one component to the other is rarely a bottleneck.

However, if we are talking about distributed systems where communication is carried over the network, this can become a problem. The time it takes to propagate the request and later response through the network can be greater than the time needed for the rest of the processing, therefore leaving the process in the idle state for most of the time (Fig 1), and delaying the response to the end user.

The second issue is the fact that downtime in one component can lead to downtime in other components: in case of component B not being reachable, component A will never receive the response, and therefore won't be able to finish processing any requests, making it unreachable as well.

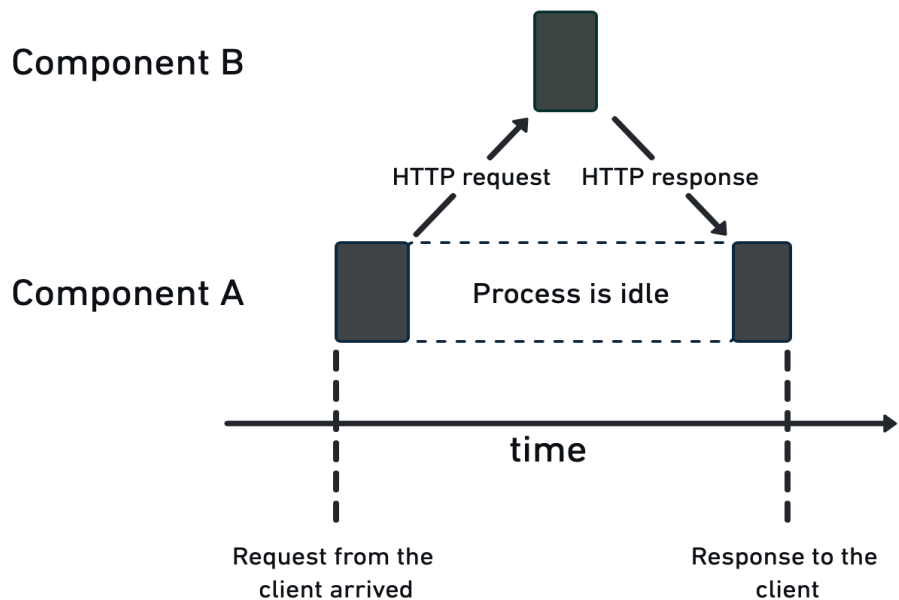

**Fig. 1.** A graph depicting synchronous communication between the components

## 2.2 Asynchronous communication

If component A can finish its task without relying on component B to complete its functionality, we can use asynchronous communication. This consists of component A sending a message to component B and proceeding with subsequent commands without waiting for the response (Fig 2). This approach will solve the problem of overall response time, as the client will get the response regardless of how long it takes to propagate messages between components. It will also avoid the propagation of errors from component B to component A, as processing will be complete regardless of whether component B is reachable.

If the task that component B performs isn't crucial, and system behavior isn't affected even though communication between two components might fail occasionally, this way of async communication is suitable.

What if component B has to correctly process each message it receives? How to keep track of messages that weren't processed? We would also need to develop some sort of a retry mechanism so that unprocessed messages can be handled.

These problems are not within the scope of any specific component in our system and are outside of the domain as well. Therefore, it would be useful to have a separate component whose sole task would be the propagation of messages between services in our application. This component needs to be fault tolerant, decoupled from implementation of any specific service that is using it, and configurable. Users should be able to define how many messages can pass through this component in any given time frame, how frequently it should try to resend messages that weren't processed, and how many times to retry before declaring that a message can not be processed.

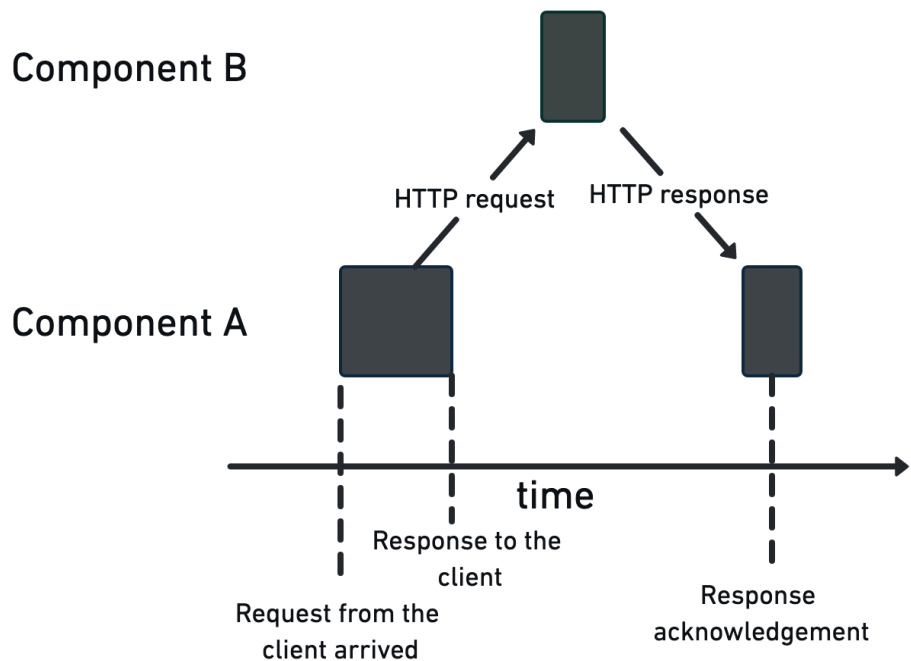

**Fig. 2.** A graph depicting asynchronous communication between the components

### 2.3 Message queues

Message queues are components utilized in distributed systems to handle asynchronous communication. Their purpose is to make communication between services fault tolerant. If the destination service is not reachable, the queue will persist the message, and retry delivering it after the specified time. This simplifies the implementation of distributed systems, such as microservice applications, as developers don't have to focus on the implementation of reliable communication.

The example used in chapters 2.1 and 2.2 to illustrate communication between components shows *point-to-point* communication [9], and as mentioned above, queues make that type of communication easier to implement. However, message queues also open up a new way of propagating information throughout the system, where a single message is delivered to multiple services, and we can dynamically change which components need to receive those messages. This model of communication is called *publisher/subscriber,* or *pub/sub* for short, and it allows for easier scaling of distributed systems [10]
.
### 2.4 Different characteristics of message queues

Depending on system requirements, we can use different types of message queues
The first thing to take into consideration is the tradeoff between message durability and system efficiency. Is it important that each message reaches its destination, even if the message queue component itself crashes unexpectedly? If so, all incoming data should be stored in the persistence layer such as the file system. which would slow down message processing but would give the queue ability to pick up where it left off after it is restarted. If we need to protect our message queue even from hardware failure, file replicas stored on different machines or a separate database need to be used, which would slow down message processing even further as it requires network calls after each message is received.

Will the queue deliver messages following the *push* or *pull* principle? Push queues are sending messages to recipients as soon as they arrive. This approach is good if we care about how quickly messages are delivered to consumers. On the other hand, this can overwhelm or even crash services if they start receiving messages faster than they can process them. Pull queues work in the opposite direction, where services that are processing data periodically ping the queue for new messages, ensuring they have enough capacity to process them.

There are a few other things that can define how message queue works, such as sending messages one by one, or in batches, and message encryption, but those fall out of the scope of this paper.

## 3  Implementation

### 3.1  Solution requirements

This paper follows the implementation of the task queue service within the constellations project. The source code for this component is open source and can be found at *https://github.com/c12s/blackhole/tree/vm/implementation_2.0*

Task queues are a subgroup of message queues focused on starting task execution on different components within the system. Requirements for the task queue service were the ability to create multiple different queues where each can be assigned a different priority, the throughput of each queue can be controlled to not overwhelm downstream services, and the number of retries and layoff period between retries can be configured.

Implementation was done in the programming language *Go, gRPC* over HTTP/2 was used for communication between services, and *protocol buffers* were used to define the Application Programming Interface (API) of task queue. All of these technologies and standards are open source.

The queue delivered messages using the *push* principle and expects acknowledgment from the recipient service that the task was successfully executed in the form of the HTTP response with status code 2xx.

### 3.2  Definition of a task

A *task* is a data structure based on which the queue knows which service it should communicate with, and how to handle if the communication or task execution is not successful. In this implementation, the task contains the following data: Unique identification, name, identification of the queue which the task should be written into, the destination of the service which should execute the task, in the form of a URL address, and which HTTP method should be used when sending the task to its destination. Additional data that can be provided with each task, if needed, includes retry policies: how many times should the queue try to execute the same task, what is the layoff period before retrying to send the same task and how long should the queue wait for the confirmation from the destination service that the task was successfully processed. Each task goes through different states, as it is being handled by the queue (Fig 3).

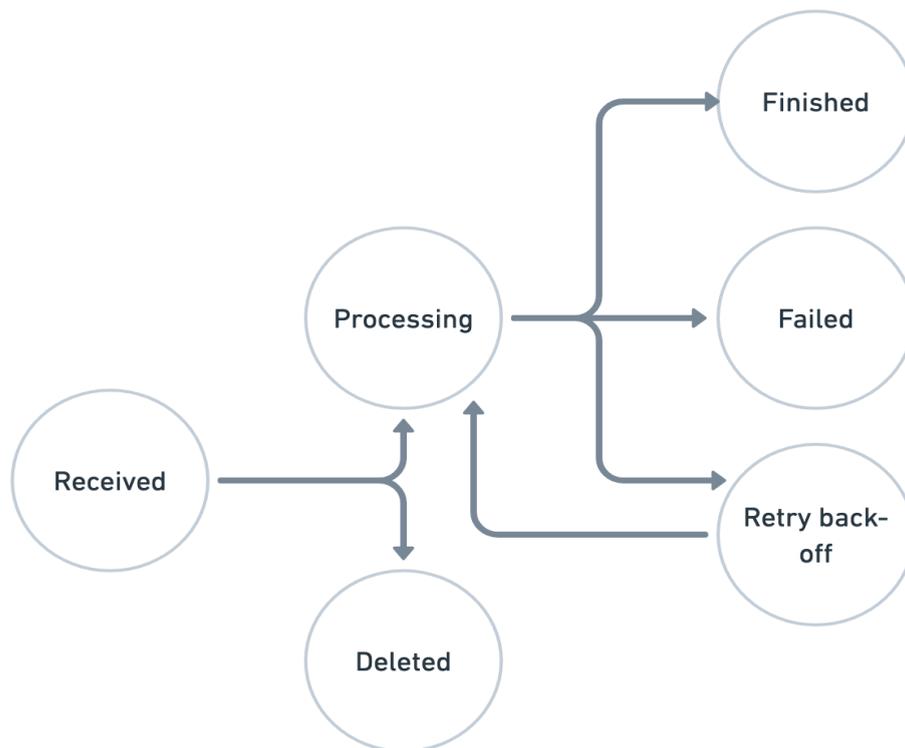

**Fig. 3.** States in which each task can be as it is being processed by the task queue

### 3.3  Definition of a task queue

A task queue is a structure used for grouping similar tasks. Up until this point the term "task queue" was used to describe the entire component in the system which handles remote execution of tasks between different services. The task queue service can have multiple different queues within it, where each of them can be used to propagate tasks (Fig 4). This chapter refers to the task queue structure within the task queue service. Metadata for each task queue includes unique UUID, unique name, the maximum number of tokens, and frequency of token regeneration (these terms are explained in chapter 3.4.).

The persistence layer used for storing tasks can be changed depending on the needs of the queue itself because this component is defined in the form of an interface. Individual queues are managed by the component named "TaskQueueManager" which is responsible for writing tasks in the corresponding queue, and forwarding them later to the workers for further processing.

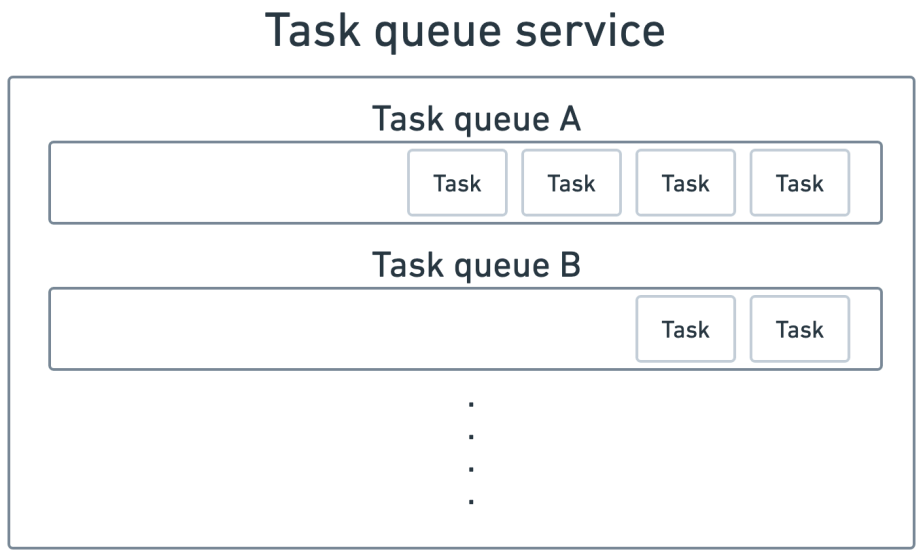

**Fig. 4.** Difference between the task queue service and individual task queues

### 3.4 Token Bucket algorithm

In systems that are used for transferring data, including message and task queues, there is a need for controlling and limiting the amount that can pass through in a certain timeframe. This is important in order to prevent other components from getting more data than they can process and exceeding their capacity. Users need a way to configure how often can tasks go from the queue to their destination to be processed

The token bucket algorithm was developed for restricting the number of packets that can pass through the telecommunication network [8]. In this paper, we will try to implement the same login for restricting the throughput of any specific queue.

Each queue is assigned a token bucked upon its creation, which consists of two parameters, the maximum number of tokens it can hold, and the frequency of generating new tokens.

When tasks reach their designated queue, they can't proceed with further processing until they are assigned a token from the bucket. With the maximum number of tokens a bucket can hold we can define how many tasks can pass at once. This parameter should reflect how many jobs can be processed in parallel by the consumer. The frequency for generating new tokens should reflect the time it takes for a single job to be processed.

Besides restricting data flow, this algorithm can be used to implicitly define the priority of each individual queue. If the frequency of new tokens in one queue is much higher than in any other, it will have the same effect as assigning that queue a higher priority.

### 3.5  Task execution

After obtaining the token, only task execution is left. Given that this operation can take some time, a standalone component within the task queue service should be implemented, so as not to burden the queue manager. This component is called a worker. It should extract the information that is contained within the task structure, construct an HTTP request based on it, and send it to the service for execution. If it fails to send the request, or the service doesn't confirm the successful task execution in the predefined timeframe, the worker checks whether it should retry sending the task, or not. If the response is no, it moves the task into the "execution failed" state, and if the answer is yes, it spends the interval predefined for waiting before attempting to send it again. These tasks don't need to wait for the token again. If we would enforce a rule where tasks need to get a token before they are sent again to their end destination, it would not be possible to guarantee the backoff time between retries that is specified within the task definition.

To achieve parallelism between the component that manages the task queue and the workers, they are executed in separate routines. Instead of creating and destroying new routines for each incoming job, it is more efficient to have a set of routines that will work non-stop - a worker pool, and to send them tasks for execution. Based on the needs of the system and available resources, we can change the number of workers that will be available.

## 4  Results and Conclusion

In this paper we implemented a dedicated software component that makes communication between distributed services reliable, fault-tolerant, and easily configurable. We defined the scope of the problem and explained the context in which the task queue would be used, as a separate service in the existing microservice platform for managing computer resources. After implementing and testing the solution we came to the conclusion that the token bucket algorithm can be used to efficiently limit the amount of messages going through the queue and also define queue priorities.

In order to see how efficient the queue is, and how implementation changes might affect it, we created a testing version of the task queue. It creates a set number of queues prefilled with generated tasks, and we measured how long it takes before all tasks have transitioned to the "Finished" state.

After defining a framework for measuring the throughput of tasks through the task queue, we experimented with different persistence layers to see how will they affect it. The conclusion we reached during these tests is that the limiting factor in task queue efficiency is the number of parallel processes/routines that can work on processing individual tasks. Writing data to persistent storage like an SSD takes considerably more time (35-40 microseconds) than writing to DRAM (100 nanoseconds) [11], but none of those tasks are bottlenecks when each worker has to deliver the message over the network and wait for the response. Choosing a technology that can easily create new processes which are efficient and lightweight, like the Go routines of Erlang processes is more important than optimizing for the time it takes to store data about each task.

### 4.1 Acknowledgment

Funded by the European Union (TaRDIS, 101093006). Views and opinions expressed are however those of the author(s) only and do not necessarily reflect those of the European Union. Neither the European Union nor the granting authority can be held responsible for them.